\documentclass[aps,prd,preprintnumbers,showpacs,showkeys,nofootinbib,superscriptaddress,fleqn,floatfix,tightenlines,10pt]{revtex4-2} 
\usepackage{amsmath,amsfonts,amssymb,amscd,amsxtra,amsthm}
\usepackage{dsfont}
\usepackage{graphicx}  
\usepackage{epstopdf}
\usepackage{dcolumn}  
\usepackage{bm}          
\usepackage{slashed}
\usepackage[utf8]{inputenc}
\usepackage{hyperref}
\usepackage{booktabs}
\usepackage[normalem]{ulem} 
\usepackage[dvipsnames]{xcolor} 
\usepackage{enumitem}
\usepackage{array}
\usepackage{slashed}
\usepackage{tikz}
\usepackage{float}
\usepackage{multirow}
\usepackage{booktabs} 
\usepackage{slashed}

\renewcommand\sout{\bgroup \color{red} \ULdepth=-.5ex \ULset}

\begin{document}
\preprint{INHA-NTG-06/2025}
\title{Triple-strangeness hidden-charm pentaquarks}
\author{Samson Clymton}
\email[E-mail: ]{samson.clymton@apctp.org}
\affiliation{Asia Pacific Center for Theoretical Physics (APCTP),
  Pohang, Gyeongbuk 37673, Republic of Korea} 

\author{Hyun-Chul Kim}
\email[E-mail: ]{hchkim@inha.ac.kr}
\affiliation{Department of Physics, Inha University,
  Incheon 22212, Republic of Korea}
\affiliation{Physics Research Institute, Inha University, Incheon
  22212, Republic of Korea} 
\affiliation{School of Physics, Korea Institute for Advanced Study 
  (KIAS), Seoul 02455, Republic of Korea}

\author{Terry Mart}
\email[E-mail: ]{terry.mart@sci.ui.ac.id}
\affiliation{Departemen Fisika, FMIPA, Universitas Indonesia, Depok
  16424, Indonesia} 
\date{\today}

\begin{abstract}
We investigate the possible existence of triple-strangeness
hidden-charm pentaquark states in the off-shell 
coupled-channel formalism. The open-charm meson-baryon 
$\bar{D}_s\Omega_c$, $\bar{D}_s\Omega_c^*$, $\bar{D}_s^*\Omega_c$, 
and $\bar{D}_s^*\Omega_c^*$ channels are considered, together with the 
hidden-charm $J/\psi\Omega$ channel. The two-body
kernel Feynman amplitudes are constructed from an effective Lagrangian
based on hidden local symmetry and heavy-quark spin symmetry. The
coupled Blankenbecler-Sugar equation is solved in the partial-wave 
helicity basis. We observe two triple-strangeness hidden-charm
pentaquark states: $P_{c\bar{c}sss}(4787)$ and
$P_{c\bar{c}sss}(4841)$, both with $J^P=1/2^-$. The
$P_{c\bar{c}sss}(4787)$ couples dominantly to the
$\bar{D}_s^*\Omega_c$ and $\bar{D}_s^*\Omega_c^*$ channels, while the 
$P_{c\bar{c}sss}(4841)$ couples almost exclusively to the
$\bar{D}_s^*\Omega_c^*$ channel. The total transition cross sections of
$\bar{D}_s^{(\ast)}\Omega_c^{(\ast)}\to J/\psi\,\Omega$ indicate that
the $P_{c\bar{c}sss}(4787)$ is clearly visible in the $J/\psi\,\Omega$
invariant mass spectrum, whereas the $P_{c\bar{c}sss}(4841)$ is
obscured by cusp structures and background contributions. 
\end{abstract}

\maketitle

\section{Introduction}
The LHCb Collaboration reported four hidden-charm pentaquark
states: $P_{c\bar{c}}(4312)$ with the decay width
$\Gamma_{P_{c\bar{c}}\to J/\psi\, p} = (10\pm 5)$~MeV,
$P_{c\bar{c}}(4380)$ with $\Gamma_{P_{c\bar{c}}\to J/\psi\, p} = 
(210\pm 90)$~MeV, $P_{c\bar{c}}(4440)$ with $\Gamma_{P_{c\bar{c}}\to
  J/\psi\, p} = (21_{-11}^{+10})$~MeV, and $P_{c\bar{c}}(4457)$ with 
$\Gamma_{P_{c\bar{c}}\to J/\psi\, p} =
(6.4_{-2.8}^{+6.0})$~MeV~\cite{LHCb:2015yax, LHCb:2019kea, LHCb:2021chn}.  They were all discovered by the LHCb Collaboration
and have not been confirmed by any other experiments. For example, the
GlueX Collaboration did not observe any signal for hidden-charm 
pentaquarks~\cite{GlueX:2019mkq}. Interestingly, these four pentaquark 
states lie below the corresponding meson--baryon thresholds. For
instance, $P_{c\bar{c}}(4312)$ and $P_{c\bar{c}}(4380)$ are located
below the $\bar{D}\Sigma_c$ and $\bar{D}\Sigma_c^*$ thresholds,
respectively, while both $P_{c\bar{c}}(4440)$ and $P_{c\bar{c}}(4457)$ are
positioned below the $\bar{D}^*\Sigma_c$ threshold. 
Subsequently, the existence of two
strangeness hidden-charm pentaquark states was reported by the LHCb  
Collaboration:
$P_{c\bar{c}s}(4338)$~\cite{LHCb:2022ogu} and
$P_{c\bar{c}s}(4459)$~\cite{LHCb:2020jpq}. 
The Belle Collaboration appears to have confirmed the existence of
$P_{c\bar{c}s}(4459)$ but reported a slightly larger mass,
$M_{P_{c\bar{c}s}} = (4471.7 \pm 4.8 \pm 0.6)$~MeV/$c^2$, and a decay
width of $\Gamma = (21.9 \pm 13.1 \pm
2.7)$~MeV~\cite{Belle:2025pey}. However, it is not clear whether 
$P_{c\bar{c}s}(4472)$ observed by the Belle experiment can be identified
as $P_{c\bar{c}s}(4459)$. For example, Clymton et
al.~\cite{Clymton:2025hez} suggested theoretically that
$P_{c\bar{c}s}(4472)$ should be distinguished from
$P_{c\bar{c}s}(4459)$. $P_{c\bar{c}s}(4459)$ and
$P_{c\bar{c}s}(4472)$ are found below the $\bar{D}^*\Xi_c$ threshold,
whereas $P_{c\bar{c}s}(4338)$ lies just about 2~MeV above the
$\bar{D}\Xi_c$ threshold. The fact that almost all observed
hidden-charm pentaquark states are located below the corresponding 
thresholds suggests that they are likely $S$-wave molecular states
composed of the corresponding heavy mesons and singly heavy baryons
associated with those thresholds, though there are more subtle
conditions required for a molecular state. 

These findings on hidden-charm pentaquarks have triggered significant
interest in heavy pentaquark states (see the recent
reviews~\cite{Esposito:2016noz, Chen:2016spr, Meng:2022ozq,
  Chen:2022asf, Huang:2023jec, Garcilazo:2025wkt} and references therein). Moreover, the CMS and LHCb
Collaborations have recently announced the    
decays $\Lambda_b^0 \to J/\psi\, \Xi^- K^+$ and $\Xi_b^0 \to J/\psi\,
\Xi^- \pi^+$~\cite{CMS:2024vnm, LHCb:2025lhk}, which indicates that
the existence of hidden-charm pentaquark states with strangeness
$S = -2$, $P_{c\bar{c}ss}$, may soon be found. Note that there have
already been theoretical predictions for such double-strangeness
hidden-charm pentaquarks~\cite{Wang:2020bjt, Ortega:2022uyu,
  Marse-Valera:2022khy, Roca:2024nsi, Marse-Valera:2024apc}, and even 
triple-strangeness ones~\cite{Meng:2019fan, Wang:2021hql,
  Roca:2024nsi}.  

In our previous works, we have investigated the production mechanisms
of hidden-charm pentaquark states with strangeness $S = 0$, $S = -1$,
and $S = -2$, based on an off-shell coupled-channel approach to
two-body charmed hadronic scattering~\cite{Clymton:2024fbf,
  Clymton:2025hez,  Clymton:2025zer}.
We first constructed the two-body kernel Feynman amplitudes for the
relevant channels by employing an effective Lagrangian that satisfies
heavy-quark spin--flavor symmetry, hidden local symmetry, and chiral
symmetry. We then computed the coupled-channel scattering integral
equations by inserting these kernel amplitudes into them. We have
seven different channels for the nonstrange hidden-charm pentaquarks,
nine channels for those with $S=-1$, and eleven channels for those
with $S=-2$.  

\begin{figure}[htp]
  \centering
  \includegraphics[scale=0.4]{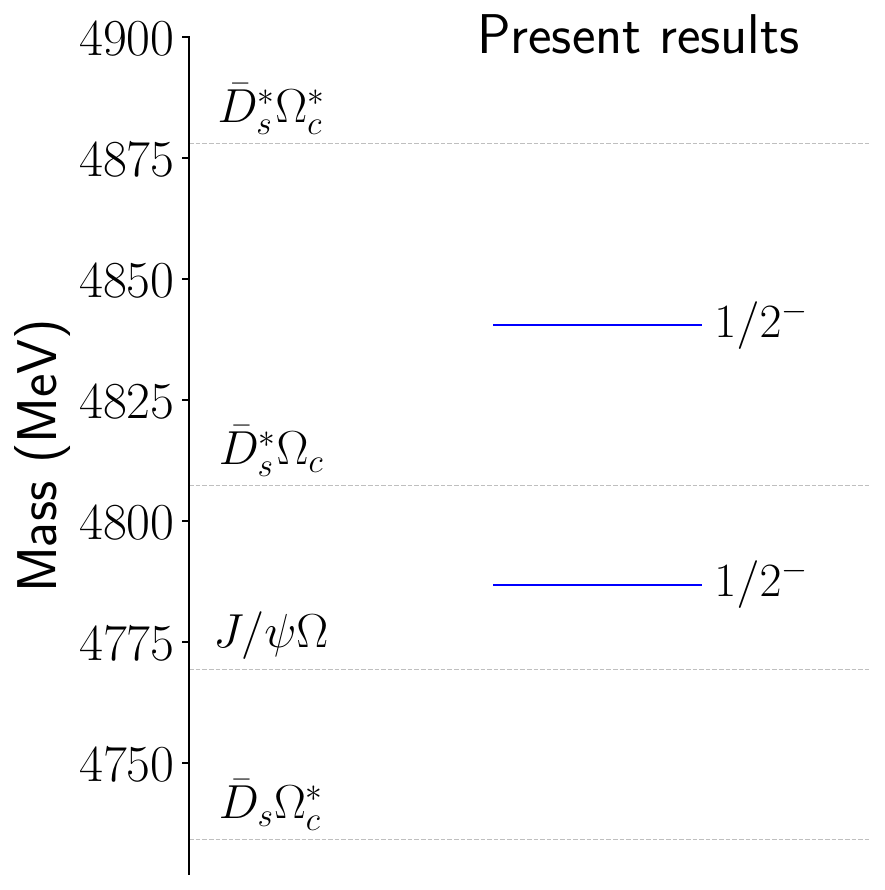}
  \caption{Predicted $P_{c\bar{c}sss}$ states from the present
    work.}  
  \label{fig:1}  
\end{figure}

The present off-shell coupled-channel formalism was successfully
applied to describing various processes~\cite{Clymton:2022jmv,
  Clymton:2023txd, Kim:2023htt, Clymton:2024pql, Kim:2025ado} in
addition to the hidden-charm pentaquarks~\cite{Clymton:2024fbf,
  Clymton:2025hez,  Clymton:2025zer}.
In this work, we extend the off-shell coupled-channel
formalism to examine how the triple-strangeness hidden-charm
pentaquark states $P_{c\bar{c}sss}$ can be generated dynamically.
To construct the transition amplitudes, we introduce five
different channels: $\bar{D}_s\Omega_c$,
$\bar{D}_s\Omega_c^*$, $J/\psi \Omega$, $\bar{D}_s^* \Omega_c$, and
$\bar{D}_s^* \Omega_c^*$. These channels mainly consist of the lowest
meson triplet ($\bm{3}$) with spins 0 and 1, and the singly heavy baryon
sextet ($\bm{6}$) with spins $1/2$ and $3/2$.  
Figure~\ref{fig:1} summarizes the predictions for the
$P_{c\bar{c}sss}$ resonances found in the present work. Two
negative-parity hidden-charm pentaquarks with $S=-3$ are obtained, as
shown in Fig.~\ref{fig:1}, both with $J^P=1/2^-$. One is located below
the $\bar{D}_s^*\Omega_c^*$ threshold, and the other lies below the
$\bar{D}_s^*\Omega_c$ threshold.

The present work is organized as follows.  
In Sec.~\ref{sec:2}, we describe the off-shell coupled-channel
formalism in the context of the $P_{c\bar{c}sss}$ states.
We first explain how the two-body kernel Feynman amplitudes with
$S=-3$ are derived from an effective Lagrangian, and then solve the 
scattering integral equations. In Sec.~\ref{sec:3}, we present and
discuss the results for the triple-strangeness hidden-charm
pentaquark states. Finally, we summarize the results, draw conclusions,
and provide an outlook. 
\section{Coupled-channel formalism\label{sec:2}} 
\begin{figure}[htbp]
  \centering
  \includegraphics[scale=1.0]{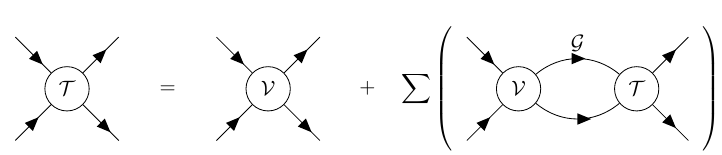}
  \caption{Graphical representation of the coupled 
          integral equation.}  
  \label{fig:2}
\end{figure}

We start with the definition of the scattering amplitude  
\begin{align}
\mathcal{S}_{fi} = \delta_{fi} - i (2\pi)^4 \delta(P_f - P_i)
  \mathcal{T}_{fi}, 
\end{align}
where $P_i$ and $P_f$ denote the total four-momenta of the initial
and final states, respectively. The transition amplitude
$\mathcal{T}_{fi}$ is constructed by using the Bethe--Salpeter equation
with the two-body Feynman kernel amplitudes:  
\begin{align}
\mathcal{T}_{fi} (p',p;s) =\, \mathcal{V}_{fi}(p',p;s) 
+ \frac{1}{(2\pi)^4}\sum_k \int d^4q \,
\mathcal{V}_{fk}(p',q;s)\,\mathcal{G}_{k}(q;s)\,\mathcal{T}_{ki}(q,p;s),  
\label{eq:BSE}
\end{align}
where $p$ and $p'$ are the relative four-momenta of the initial and
final states, and $q$ denotes the off-shell momentum of the
intermediate states in the center-of-mass (CM) frame. The variable
$s \equiv P_i^2 = P_f^2$ stands for the square of the total energy
and is one of the Mandelstam variables. The indices $i$, $f$, and $k$
represent the initial, final, and intermediate states, respectively.
Since we consider coupled channels relevant to the production of
$P_{c\bar{c}sss}$ states, the index $k$ runs over all relevant
intermediate channels. The coupled integral scattering equations
given in Eq.~\eqref{eq:BSE} are illustrated schematically in
Fig.~\ref{fig:2}.  

To simplify the calculation of the four-dimensional integral
equations, we perform a three-dimensional reduction. Among the  
various methods available, we employ the Blankenbecler--Sugar
formalism~\cite{Blankenbecler:1965gx, Aaron:1968aoz}. 
The two-body propagator is given by  
\begin{align}
  \mathcal{G}_k(q) =\;
  \delta\left(q_0-\frac{E_{k1}(\bm{q})-E_{k2}(\bm{q})}{2}\right)
  \frac{\pi}{E_{k1}(\bm{q})E_{k2}(\bm{q})}
  \frac{E_k(\bm{q})}{s-E_k^2(\bm{q})},
\label{eq:tbprop}
\end{align}
where $E_k$ is the total on-shell energy of the
intermediate state, defined as $E_k = E_{k1} + E_{k2}$, and $\bm{q}$
denotes the three-momentum of the intermediate state. Note that the
spinor contributions from the meson--baryon propagator $G_k$ have been 
absorbed into the matrix elements of $\mathcal{V}$ and $\mathcal{T}$.  
Using Eq.~\eqref{eq:tbprop}, we obtain the following coupled integral
equations:      
\begin{align}
  \mathcal{T}_{fi} (\bm{p}',\bm{p}) =\, \mathcal{V}_{fi}
  (\bm{p}',\bm{p}) 
  +\frac{1}{(2\pi)^3}\sum_k\int \frac{d^3q}{2E_{k1}(\bm{q})E_{k2}
  (\bm{q})} \mathcal{V}_{fk}(\bm{p}',\bm{q})\frac{E_k
  (\bm{q})}{s-E_k^2(\bm{q})+i\varepsilon} 
  \mathcal{T}_{ki}(\bm{q},\bm{p}),
  \label{eq:BS-3d}
\end{align}
where $\bm{p}$ and $\bm{p}'$ denote the relative three-momenta of
the initial and final states, respectively, in the CM frame.

\begin{figure}[htbp]
  \centering
  \includegraphics[scale=0.35]{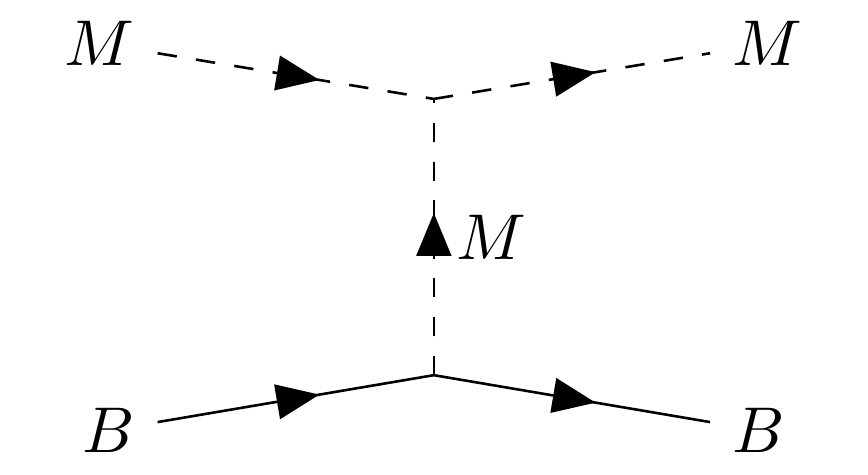}
  \caption{$t$-channel meson-exchange diagrams. $M$ and $B$ denote a 
  meson and a baryon involved in the process, respectively.}  
  \label{fig:3}
\end{figure}
To generate the $P_{c\bar{c}sss}$ states, we construct two-body
coupled channels by combining the charmed meson triplet ($\bm{3}_0$
and $\bm{3}_1$) with the singly charmed baryon sextet ($\bm{6}_{1/2}$
and $\bm{6}_{3/2}$), while fixing the strangeness to
$S=-3$. We also include the $J/\psi \,\Omega$ channel, given that both
the $P_{c\bar{c}}$ and $P_{c\bar{c}s}$ states were observed in the
invariant mass spectra of $J/\psi N$ and $J/\psi \Lambda$,
respectively.  
Thus, we consider five distinct channels: $\bar{D}_s\,\Omega_c$, 
$\bar{D}_s\,\Omega_c^*$, $J/\psi\,\Omega$,
$\bar{D}_s^*\,\Omega_c$, and $\bar{D}_s^*\,\Omega_c^*$.  
The two-body kernel Feynman amplitudes are derived from the
one-meson-exchange diagrams shown in Fig.~\ref{fig:3}.  
We exclude $s$-channel pole diagrams, since our aim is to investigate
how the $P_{c\bar{c}sss}$ states can be dynamically generated.  
We also omit the $u$-channel diagrams, as they involve doubly charmed
baryons with masses of approximately 3.5~GeV, which makes their
contributions significantly suppressed compared to the
$t$-channel contributions.

The vertex interactions are determined by using an effective
Lagrangian that satisfies heavy-quark spin symmetry, hidden local
symmetry, and flavor SU(3) symmetry~\cite{Casalbuoni:1996pg}, which
are given by  
\begin{align}
  \mathcal{L}_{PP\mathbb{V}} &= -i\frac{\beta g_V}{\sqrt{2}}\,
  P^{\dagger}_a  \overleftrightarrow{\partial_\mu} P_b\,
   \mathbb{V}^\mu_{ba} +i\frac{\beta g_V}{\sqrt{2}}\,P'^{\dagger}_a
                               \overleftrightarrow{\partial_\mu}
                               P'_b\, \mathbb{V}^\mu_{ab},
  \label{eq:5} \\ 
    \mathcal{L}_{PP\sigma} &= -2g_\sigma M P^\dagger_a \sigma P_a
  -2g_\sigma M P'^\dagger_a\sigma P'_a , \label{eq:6}\\     
    \mathcal{L}_{P^*P^*\mathbb{P}} &= -\frac{g}{f_\pi}
  \epsilon^{\mu\nu\alpha\beta}P^{*\dagger}_{a\nu}\,
   \overleftrightarrow{\partial_\mu}\,
P^*_{b\beta}\partial_\alpha \mathbb{P}_{ba} -\frac{g}{f_\pi}
  \epsilon^{\mu\nu\alpha\beta}  P'^{*\dagger}_{a\nu}\,
 \overleftrightarrow{\partial_\mu}\,
 P'^*_{b\beta}\partial_\alpha
                                     \mathbb{P}_{ab} , \label{eq:7} \\     
    \mathcal{L}_{P^*P^*\mathbb{V}} & = i\frac{\beta g_V}{\sqrt{2}} \,
 P^{*\dagger}_{a\nu}  \overleftrightarrow{\partial_\mu}
 P^{*\nu}_b \mathbb{V}_{ba}^\mu + i2\sqrt{2} \lambda g_VM^*
 P^{*\dagger}_{a\mu}  P^*_{b\nu}\mathbb{V}_{ba}^{\mu\nu}\cr
  &\;\;\;\;-i\frac{\beta g_V}{\sqrt{2}} \,P'^{*\dagger}_{a\nu}
    \overleftrightarrow{\partial_\mu} P'^{*\nu}_b
    \mathbb{V}_{ab}^\mu-i2\sqrt{2}\lambda g_VM^*
    P'^{*\dagger}_{a\mu} P'^*_{b\nu}\mathbb{V}_{ab}^{\mu\nu} ,
  \label{eq:8}\\
  \mathcal{L}_{P^*P^*\sigma} &= 2g_\sigma M^* P^{*\dagger}_{a\mu}
\sigma P^{*\mu}_a+2g_\sigma M^*
 P'^{*\dagger}_{a\mu}\sigma P'^{*\mu}_a , \label{eq:9}\\    
    \mathcal{L}_{P^*P\mathbb{P}} &= -\frac{2g}{f_\pi}  \sqrt{MM^*}\,
      \left( P^{\dagger}_a P^*_{b\mu}+P^{*\dagger}_{a\mu} P_b\right)\,
 \partial^\mu \mathbb{P}_{ba}+\frac{2g}{f_\pi} \sqrt{MM^*}\,
    \left(P'^{\dagger}_a P'^*_{b\mu}+P'^{*\dagger}_{a\mu} P'_b\right)\,
  \partial^\mu \mathbb{P}_{ab}, \label{eq:10} \\
  \mathcal{L}_{P^*P\mathbb{V}} &= -i\sqrt{2}\lambda g_V\,
\epsilon^{\beta\alpha\mu\nu} \left(P^{\dagger}_a
\overleftrightarrow{\partial_\beta} P^*_{b\alpha} +
P^{*\dagger}_{a\alpha}  \overleftrightarrow{\partial_\beta}
  P_b\right)\,\left(\partial_\mu\mathbb{V}_{\nu}\right)_{ba}\cr 
 &\;\;\;\;-i\sqrt{2}\lambda g_V\, \epsilon^{\beta\alpha\mu\nu}
   \left(P'^{\dagger}_a\overleftrightarrow{\partial_\beta}
   P'^*_{b\alpha}+P'^{*\dagger}_{a\alpha}
   \overleftrightarrow{\partial_\beta}P'_b\right)\,
   \left(\partial_\mu\mathbb{V}_{\nu}\right)_{ab},
\label{eq:11}
\end{align}
where $\overleftrightarrow{\partial} =
\overrightarrow{\partial}-\overleftarrow{\partial}$.  
Here, $\sigma$ represents the lowest-lying isoscalar-scalar meson,
i.e., $f_0(500)$.  Heavy mesons and anti-heavy mesons $P^{(*)}$ and
$P'^{(*)}$ are defined as  
\begin{align}
  P = \left(D^0,D^+,D_s^+\right), \hspace{0.5 cm}
  P^*_\mu =\left(D^{*0}_\mu,D^{*+}_\mu,D_{s\mu}^{*+}\right),
  \hspace{0.5 cm} P' =(\bar{D}^0,\,D^-,\,D_s^-),
  \hspace{0.5 cm} P'^*_\mu
  =(\bar{D}^{*0}_\mu,\,D^{*-}_\mu,\,D^{*-}_{s\mu}), 
\end{align}
while the light pseudoscalar and vector mesons
are expressed as  
\begin{align}
    \mathbb{P} = 
    \begin{pmatrix}
        \frac{1}{\sqrt{2}} \pi^0+\frac{1}{\sqrt{6}}\eta & \pi^+ & K^+\\
        \pi^- & -\frac{1}{\sqrt{2}} \pi^0+\frac{1}{\sqrt{6}}\eta & K^0\\
        K^- & \bar{K}^0 & -\frac{2}{\sqrt{6}}\eta
    \end{pmatrix},\;\;\;\;
    \mathbb{V}_\mu = \begin{pmatrix}
        \frac{1}{\sqrt{2}} \rho^0_\mu+\frac{1}{\sqrt{2}}\omega_\mu &
        \rho_\mu^+ & K_\mu^{*+}\\
        \rho_\mu^- & -\frac{1}{\sqrt{2}} \rho_\mu^0+\frac{1}{\sqrt{2}}
        \omega_\mu & K_\mu^{*0} \\
        K_\mu^{*-} & \bar{K}^{*0}_\mu & \phi_\mu
    \end{pmatrix}.
\end{align}

Since we have already determined all the coupling constants pertinent
also to this work, we will briefly mention their values. The value of
$g_V$ in Eqs.~\eqref{eq:5},~\eqref{eq:8}, and~\eqref{eq:11} is obtained
from the Kawarabayashi--Suzuki--Riazuddin--Fayyazuddin (KSRF)
relation~\cite{Kawarabayashi:1966kd, Riazuddin:1966sw} 
$g_V = m_\rho / f_\pi \approx 5.8$, where $f_\pi = 132~\mathrm{MeV}$. 
The value of $\beta$ in Eqs. ~\eqref{eq:5} and~\eqref{eq:8}
is taken to be $\beta\approx 0.9$ by employing the vector-meson
dominance in heavy-meson radiative decays. 
We fix $g = 0.59 \pm 0.07 \pm 0.01$ in Eqs.~\eqref{eq:7}
and~\eqref{eq:10} by using the experimental data on the full width of
$D^{*+}$~\cite{Isola:2003fh}. The value of $\lambda$ in
Eq.~\eqref{eq:11} is given as $\lambda = -0.56~\mathrm{GeV}^{-1}$,
extracted from light-cone sum rules and lattice QCD.   
Note that the sign convention for $\lambda$ is different from that in
Ref.~\cite{Isola:2003fh}, since we use the same phase for heavy vector 
mesons as in Ref.~\cite{Casalbuoni:1996pg}.  
For the $\sigma$ meson in Eq.~\eqref{eq:6} and~\eqref{eq:9}, the
coupling constant $g_\sigma$ is determined from the analysis of the
$2\pi$ transition of $D_s(1^+)$ in Ref.~\cite{Bardeen:2003kt}: 
$g_\sigma = g_\pi/(2\sqrt{6})$ with $g_\pi = 3.73$.

Regarding the heavy baryon effective Lagrangian, we follow the framework 
developed by Ref.~\cite{Liu:2011xc}, which incorporates a more
comprehensive Lagrangian formulation as outlined in
Ref.~\cite{Yan:1992gz}. The baryonic interaction vertices within the
tree-level meson-exchange diagrams are given by the following
effective Lagrangian:  
\begin{align}
    \mathcal{L}_{B_6B_6\mathbb{P}}&= i\frac{g_1}{2f_\pi M_6}\bar{B}_{6}\gamma_5\left(\gamma^\alpha\gamma^\beta-g^{\alpha\beta}\right)\overleftrightarrow{\partial_\alpha}\partial_\beta\mathbb{P} B_{6} ,\\
    \mathcal{L}_{B_6B_6\mathbb{V}}&= -i\frac{\beta_6 g_V}{2\sqrt{2}M_6}\left(\bar{B}_{6}\overleftrightarrow{\partial_\alpha}\mathbb{V}^\alpha B_{6}\right)-\frac{i\lambda_6g_V}{3\sqrt{2}}\left(\bar{B}_{6}\gamma_\mu\gamma_\nu \mathbb{V}^{\mu\nu}B_{6}\right) ,\\
    \mathcal{L}_{B_6B_6\sigma}&= -l_6\left(\bar{B}_{6}\sigma B_6\right) ,\\
    \mathcal{L}_{B_6^*B_6^*\mathbb{P}}&= \frac{3g_1}{4f_\pi M_6^*}\epsilon^{\mu\nu\alpha\beta}\left(\bar{B}_{6\mu}^*\overleftrightarrow{\partial_\nu}\partial_\alpha\mathbb{P} B_{6\beta}^*\right) ,\\
    \mathcal{L}_{B_6^*B_6^*\mathbb{V}}&= i\frac{\beta_6 g_V}{2\sqrt{2}M_6^*}\left(\bar{B}_{6\mu}^*\overleftrightarrow{\partial_\alpha}\mathbb{V}^\alpha B_{6}^{*\mu}\right)+\frac{i\lambda_6g_V}{\sqrt{2}} \left(\bar{B}^*_{6\mu} \mathbb{V}^{\mu\nu}B^*_{6\nu}\right) ,\\
    \mathcal{L}_{B_6^*B_6^*\sigma}&= l_6\left(\bar{B}^*_{6\mu}\sigma B^{*\mu}_6\right) ,\\
    \mathcal{L}_{B_6B_6^*\mathbb{P}}&= \frac{g_1}{4f_\pi}\sqrt{\frac{3}{M_6^*M_6}}\epsilon^{\mu\nu\alpha\beta}\left[\left(\bar{B}_{6}\gamma_5\gamma_\mu\overleftrightarrow{\partial_\nu} \partial_\alpha\mathbb{P} B_{6\beta}^*\right)+\left(\bar{B}_{6\mu}^*\gamma_5\gamma_\nu\overleftrightarrow{\partial_\alpha} \partial_\beta\mathbb{P}B_6 \right)\right] ,\\
    \mathcal{L}_{B_6B_6^*\mathbb{V}}&= \frac{i\lambda_6g_V}{\sqrt{6}}\left[\bar{B}_{6}\gamma_5\left(\gamma_\mu +\frac{i\overleftrightarrow{\partial_\mu}}{2\sqrt{M_6^*M_6}}\right) \mathbb{V}^{\mu\nu}B^*_{6\nu}+\bar{B}_{6\mu}^*\gamma_5\left(\gamma_\nu -\frac{i\overleftrightarrow{\partial_\nu}}{2\sqrt{M_6^*M_6}}\right) \mathbb{V}^{\mu\nu}B_{6}\right] ,
\end{align}
with heavy baryon fields expressed as
\begin{align}
    B_6 =
    \begin{pmatrix}
        \Sigma_c^{++} & \frac{1}{\sqrt{2}}\Sigma_c^+ & \frac{1}{\sqrt{2}}\Xi{'}_c^+\\
        \frac{1}{\sqrt{2}}\Sigma_c^+ & \Sigma_c^0 & \frac{1}{\sqrt{2}}\Xi{'}_c^0\\
        \frac{1}{\sqrt{2}}\Xi{'}_c^+ & \frac{1}{\sqrt{2}}\Xi{'}_c^0 & \Omega_c^0
    \end{pmatrix},\;\;
    B_6^* =
    \begin{pmatrix}
        \Sigma_c^{*++} & \frac{1}{\sqrt{2}}\Sigma_c^{*+} & \frac{1}{\sqrt{2}}\Xi_c^{*+}\\
        \frac{1}{\sqrt{2}}\Sigma_c^{*+} & \Sigma_c^{*0} & \frac{1}{\sqrt{2}}\Xi_c^{*0}\\
        \frac{1}{\sqrt{2}}\Xi_c^{*+} & \frac{1}{\sqrt{2}}\Xi_c^{*0} & \Omega_c^{*0}
    \end{pmatrix}.
\end{align}
The symbol $B_\mu$ denotes the spin 3/2 Rarita-Schwinger field, which
is subject to the constraints 
\begin{align}
  p^\mu B_\mu = 0 \hspace{0.5 cm}{\rm and}\hspace{0.5 cm}
  \gamma^\mu B_\mu = 0 .
\end{align}
The coupling constants for the baryonic vertices are fixed 
as follows~\cite{Liu:2011xc,Chen:2019asm}: $\beta_6=-12/g_V$, 
$\lambda_6=-3.31\,\mathrm{GeV}^{-1}$, $g_1=0.942$ and $l_6=6.2$. 
The sign conventions we implement are in line with those 
given in Refs.~\cite{Chen:2019asm, Dong:2021juy}.  

We include only the $J/\psi \Omega$ channel, considering the fact
that the $P_{c\bar{c}}$ and $P_{c\bar{c}s}$ can decay into the $J/\psi
N$ and $J/\psi \Lambda$ states, respectively. We expect that the
$P_{c\bar{c}sss}$ states may decay into the $J/\psi \Omega$ state, if
they exist experimentally. The interaction between heavy mesons and
charmonium is expressed as the following effective
Lagrangian~\cite{Colangelo:2003sa}: 
\begin{align}
  \mathcal{L}_{PPJ/\psi} &= -ig_\psi M\sqrt{m_{J}}
  \left(J/\psi^\mu P^\dagger\overleftrightarrow{\partial_\mu}
  P{'}^{\dagger}\right) + \mathrm{h.c.},\\
  \mathcal{L}_{P^*PJ/\psi} &= ig_\psi\sqrt{\frac{MM^*}{m_{J}}}
  \epsilon^{\mu\nu\alpha\beta} \partial_\mu J/\psi_\nu
 \left(P^\dagger\overleftrightarrow{\partial_\alpha}
P^*{'}^\dagger_\beta+P_{\beta}^{*\dagger}
  \overleftrightarrow{\partial_\alpha}P{'}^{\dagger}\right)
   +\mathrm{h.c.},\\
  \mathcal{L}_{P^*P^*J/\psi} &= ig_\psi M^*\sqrt{m_J}
  (g^{\mu\nu}g^{\alpha\beta}-g^{\mu\alpha}g^{\nu\beta}
 +g^{\mu\beta}g^{\nu\alpha}) \left(J/\psi_\mu
   P_{\nu}^{*\dagger}\overleftrightarrow{\partial_\alpha}
   P^*{'}^\dagger_\beta\right)+\mathrm{h.c.}.
\end{align}
Since no experimental data exist for the $J/\psi \to
D\bar{D}$ decay, we follow Ref.~\cite{Shimizu:2017xrg}, where 
the coupling constant $g_\psi$ was estimated to be $g_\psi =
0.679\,\mathrm{GeV}^{-3/2}$.    

As for the coupling of baryon decuplet with baryon sextet, we employ
the following effective Lagrangian with heavy quark spin symmetry: 
\begin{align}
 \mathcal{L}_{\Omega_c D_s\Omega} &= g_{I}\sqrt{\frac{M_{D_s}}{3}}\,\bar{\Omega}_c\,\left(\gamma_\mu -\frac{i\overleftrightarrow{\partial_\mu}}{2M_{D_s}}\right) D_s \Omega^\mu+\mathrm{h.c.},\\
 \mathcal{L}_{\Omega_c D_s^*\Omega} &= g_{I}\sqrt{\frac{M_{D_s^*}}{3}}\,\bar{\Omega}_c\,\gamma_5\left(\gamma_\mu +\frac{i\overleftrightarrow{\partial_\mu}}{2M_{D_s^*}}\right)\gamma_\nu D_s^{*\nu} \Omega^\mu+\mathrm{h.c.},\\
 \mathcal{L}_{\Omega_c^* D_s\Omega} &= -g_{I}\sqrt{M_{D_s}}\,\bar{\Omega}_{c\mu}^*\gamma_5 D_s \Omega^\mu+\mathrm{h.c.},\\
 \mathcal{L}_{\Omega_c^* D_s\Omega} &= g_{I}\sqrt{M_{D_s}}\,\bar{\Omega}_{c\mu}^*\gamma_\nu D_s^{*\nu} \Omega^\mu+\mathrm{h.c.}.
\end{align}
The coupling constant $g_I$ is of the same order as the
coupling constant for the baryon octet-baryon sextet-meson triplet
vertex used in our previous work~\cite{Clymton:2025zer}. We therefore
use the value $g_I = 1\,\mathrm{GeV}^{-1/2}$, which renders the 
$J/\psi\Omega$ channel with negligible effects on the coupled-channel
dynamics.  

\begin{table}[htbp]
  \caption{\label{tab:1}Values of the IS factors 
          for the corresponding $t$-channel diagrams for
          the given reactions.  
        } 
   \renewcommand{\arraystretch}{1.2}
  \begin{ruledtabular}
  \centering\begin{tabular}{lcr}
   \multirow{2}{*}{Reactions} & \multirow{2}{*}{Exchange particles} & 
   \multirow{2}{*}{IS} \\
   & &
   \\
   \hline\\[-2.5ex]
     $\bar{D}_s\Omega_c\to\bar{D}_s\Omega_c$ 
     & $\phi$  & $1$ \\
     & $\sigma$  & $1$ \\
     $\bar{D}_s\Omega_c\to\bar{D}_s\Omega_c^*$ 
     & $\phi$  & $1$ \\
     $\bar{D}_s\Omega_c\to J/\psi\Omega$ 
     & $\bar{D}_s$, $\bar{D}_s^*$  & $1$ \\
     $\bar{D}_s\Omega_c\to\bar{D}_s^*\Omega_c$ 
     & $\eta$  & $\frac{1}{3}\sqrt{2}$ \\
     & $\phi$  & $1$ \\
     $\bar{D}_s\Omega_c\to\bar{D}_s^*\Omega_c^*$ 
     & $\eta$  & $\frac{1}{3}\sqrt{2}$ \\
     & $\phi$  & $1$ \\
     $\bar{D}_s\Omega_c^*\to\bar{D}_s\Omega_c^*$ 
     & $\phi$  & $1$ \\
     & $\sigma$  & $1$ \\
     $\bar{D}_s\Omega_c^*\to J/\psi\Omega$ 
     & $\bar{D}_s$, $\bar{D}_s^*$  & $1$ \\
     $\bar{D}_s\Omega_c^*\to\bar{D}_s^*\Omega_c$ 
     & $\eta$  & $\frac{1}{3}\sqrt{2}$ \\
     & $\phi$  & $1$ \\
     $\bar{D}_s\Omega_c^*\to\bar{D}_s^*\Omega_c^*$ 
     & $\eta$  & $\frac{1}{3}\sqrt{2}$ \\
     & $\phi$  & $1$ \\
     $J/\psi\Omega\to\bar{D}_s^*\Omega_c$ 
     & $\bar{D}_s$, $\bar{D}_s^*$  & $1$ \\
     $J/\psi\Omega\to\bar{D}_s^*\Omega_c^*$ 
     & $\bar{D}_s$, $\bar{D}_s^*$  & $1$ \\
     $\bar{D}_s^*\Omega_c\to\bar{D}_s^*\Omega_c$ 
     & $\eta$  & $\frac{1}{3}\sqrt{2}$ \\
     & $\phi$  & $1$ \\
     & $\sigma$  & $1$ \\
     $\bar{D}_s^*\Omega_c\to\bar{D}_s^*\Omega_c^*$ 
     & $\eta$  & $\frac{1}{3}\sqrt{2}$ \\
     & $\phi$  & $1$ \\
     $\bar{D}_s^*\Omega_c^*\to\bar{D}_s^*\Omega_c^*$ 
     & $\eta$  & $\frac{1}{3}\sqrt{2}$ \\
     & $\phi$  & $1$ \\
     & $\sigma$  & $1$ \\
  \end{tabular}
    \end{ruledtabular}
\end{table}
The two-body kernel Feynman amplitude for a one-meson exchange diagram 
can be generically expressed as    
\begin{align}
  \mathcal{A}_{\lambda'_1\lambda'_2,\lambda_1\lambda_2}
  = \mathrm{IS} \,F^2(q^2)\,\Gamma_{\lambda'_1\lambda'_2}(p'_1,p'_2)
  \mathcal P(q)\Gamma_{\lambda_1\lambda_2}(p_1,p_2) ,
\end{align}
where $\lambda_i$ and $p_i$ are respectively the helicity and
momentum of each particle participating in the process, while $q$
stands for the momentum transfer. The IS factor contains the SU(3) 
Clebsch-Gordan coefficient and isospin factor, of which the values for each
exchange diagram listed in Table~\ref{tab:1}. The vertex functions
$\Gamma$ can be directly determined by the effective Lagrangian.  
We derive the propagators $\mathcal{P}(q)$ for spin-0 and spin-1
mesons as    
\begin{align}
  \mathcal{P}(q) &= \frac{1}{q^2-m^2},\;\;\;
  \mathcal{P}_{\mu\nu}(q) = \frac{1}{q^2-m^2}
  \left(-g_{\mu\nu}+\frac{q_\mu q_\nu}{m^2}\right).
\end{align}
Since we take the finite value of the heavy-quark mass, we use the
same form for the heavy mesons as for the light mesons. 

We introduce a form factor at each vertex, since hadrons have finite
sizes. We employ the following form of the form
factor~\cite{Kim:1994ce} 
\begin{align}
F(q^2) = \left(\frac{n\Lambda^2-m^2}
{n\Lambda^2-q^2}
\right)^n,
\label{eq:39}
\end{align}
where the value of $n$ varies depending on the momentum power present
in the vertex function $\Gamma$. This parametrization has an essential
merit, since we do not need to modify $\Lambda$ when $n$ changes. Note
that as $n$ approaches infinity, Eq.~\eqref{eq:39} becomes a Gaussian
form. While it is difficult to fix the cutoff masses $\Lambda$ in
Eq.~\eqref{eq:39} due to the lack of experimental information, we make
use of the physical property of hadron sizes to reduce the
uncertainties associated with the cutoff masses. Investigations of the
electromagnetic form factors of singly heavy
baryons~\cite{Kim:2018nqf, Kim:2021xpp} have shown that heavy hadrons
exhibit more compact sizes than their lighter counterparts. In other
words, the cutoff mass $\Lambda$ is proportional to the inverse of the
radius of the corresponding particle, $\Lambda \sim \langle
r\rangle^{-1}$. This indicates that higher cutoff masses are
appropriate for heavy hadrons compared to light hadrons. Following
this argument, we define the reduced cutoff mass as $\Lambda_0 :=
\Lambda - m$, where $m$ denotes the mass of the exchange particle. In
the absence of experimental data on the $P_{c\bar{c}sss}$, we adopt a
common value of 700 MeV for all reduced cutoff parameters
$\Lambda_0$. Moreover, Cheng et al. proposed that the cutoff
mass can be related to $\Lambda_{\mathrm{QCD}}$~\cite{Cheng:2004ru},
i.e., $\Lambda = m+\eta \Lambda_{\mathrm{QCD}}$, where $\eta$ is of
order unity. It depends on the exchanged particle as well as on the 
external particles involved in the strong interaction vertex. The
value of $\eta$ tends to be larger as the mass of the exchange
particle increases. In addition, the corresponding cutoff masses
increase with the mass of the external particles. Although this
tendency cannot be proven mathematically, it has been shown to be
plausible phenomenologically in previous works~\cite{Clymton:2022jmv,
  Clymton:2023txd, Kim:2023htt, Clymton:2024pql, Kim:2025ado,
  Clymton:2024fbf, Clymton:2025hez, Clymton:2025zer}.

Since the strong interaction is invariant under parity transformation,
the number of transition amplitudes can be reduced by employing parity
invariance: 
\begin{align}
\mathcal{A}_{-\lambda'_1-\lambda'_2,-\lambda_1-\lambda_2} =
\eta(\eta')^{-1}
\mathcal{A}_{\lambda'_1\lambda'_2,\lambda_1\lambda_2},
\label{eq:ampi}
\end{align}
where $\eta$ and $\eta'$ are defined as
\begin{align}
\eta \equiv \eta_1\eta_2(-1)^{J-s_1-s_2}, \hspace{0.5 cm}
\eta' \equiv \eta_1'\eta_2'(-1)^{J-s_1'-s_2'}.
\end{align}
Here, $\eta_i$ and $s_i$ denote the intrinsic parity and spin of the
$i$-th particle, respectively, while $J$ represents the total angular
momentum. The relation in Eq.~\eqref{eq:ampi} significantly reduces
the computational time required for numerical calculations.

To classify the spin-parity $J^P$ of the $P_{c\bar{c}sss}$ states, we
perform the partial-wave decomposition of the $\mathcal{V}$ and
$\mathcal{T}$ matrices: 
\begin{align}
\mathcal{T}^{J(fi)}_{\lambda'\lambda} (\mathrm{p}',\mathrm{p}) =
\mathcal{V}^{J(fi)}_{\lambda'\lambda} (\mathrm{p}',\mathrm{p})+
\frac{1}{(2\pi)^3} \sum_{k,\lambda_k} \int 
\frac{\mathrm{q}^2d\mathrm{q}}{2E_{k1}E_{k2}}
\mathcal{V}^{J(fk)}_{\lambda'\lambda_k}(\mathrm{p}',\mathrm{q})
\frac{E_k}{s-E_k^2+i\varepsilon}
\mathcal{T}^{J(ki)}_{\lambda_k\lambda} (\mathrm{q},\mathrm{p}),
\label{eq:BS-1d}
\end{align}
where the helicities of the final, initial, and intermediate states
are denoted by $\lambda'=\{\lambda'_1,\lambda'_2\}$,
$\lambda=\{\lambda_1,\lambda_2\}$, and
$\lambda_k=\{\lambda_{k1},\lambda_{k2}\}$, respectively. The variables
$\mathrm{p}'$, $\mathrm{p}$, and $\mathrm{q}$ represent the magnitudes
of the corresponding three-momenta $\bm{p}'$, $\bm{p}$, and $\bm{q}$,
respectively. 

The partial-wave expansion for the kernel amplitudes
$\mathcal{V}^{J(fi)}_{\lambda'\lambda}$ is given by 
\begin{equation}
\mathcal{V}^{J(fi)}_{\lambda'\lambda}(\mathrm{p}',\mathrm{p}) =
2\pi \int d(\cos\theta)
d^{J}_{\lambda_1-\lambda_2,\lambda'_1-\lambda'_2}(\theta)
\mathcal{V}^{fi}_{\lambda'\lambda}(\mathrm{p}',\mathrm{p},\theta),
\label{eq:pwd}
\end{equation}
where $\theta$ is the scattering angle and
$d^{J}_{\lambda\lambda'}(\theta)$ denotes the reduced Wigner $D$
functions. 

Singularities arise from the two-body propagator $\mathcal{G}$ in the
coupled integral equation in Eq.~\eqref{eq:BS-1d}, which can be
isolated to handle. So, the integral equation is decomposed into the
regularized one and the singular part:  
\begin{align}
  \mathcal{T}^{fi}_{\lambda'\lambda} (\mathrm{p}',\mathrm{p}) = 
  \mathcal{V}^{fi}_{
  \lambda'\lambda} (\mathrm{p}',\mathrm{p}) + \frac{1}{(2\pi)^3}
  \sum_{k,\lambda_k}\left[\int_0^{\infty}d\mathrm{q}
  \frac{\mathrm{q}E_k}{E_{k1}E_{k2}}\frac{\mathcal{F}(\mathrm{q})
  -\mathcal{F}(\tilde{\mathrm{q}}_k)}{s-E_k^2}+ \frac{1}{2\sqrt{s}}
  \left(\ln\left|\frac{\sqrt{s}-E_k^{\mathrm{thr}}}{\sqrt{s}
  +E_k^{\mathrm{thr}}}\right|-i\pi\right)\mathcal{F}
  (\tilde{\mathrm{q}}_k)\right],
  \label{eq:BS-1d-reg}
\end{align}
where
\begin{align}
  \mathcal{F}(\mathrm{q})=\frac{1}{2}\mathrm{q}\,
  \mathcal{V}^{fk}_{\lambda'\lambda_k}(\mathrm{p}',
  \mathrm{q})\mathcal{T}^{ki}_{\lambda_k\lambda}(\mathrm{q},\mathrm{p}) ,
\end{align}
and $\tilde{\mathrm{q}}_k$ denotes the momentum $\mathrm{q}$ when
$E_{k1}+E_{k2}=\sqrt{s}$. This regularization procedure is exclusively
implemented when the total energy $\sqrt{s}$ is greater than the 
threshold energy of the $k$-th channel $E_k^{\mathrm{thr}}$. Notably,
the form factors in the kernel amplitudes $\mathcal{V}$
guarantee the unitarity of the transition amplitudes in the
high-momentum domain. 

To perform the numerical evaluation of the $\mathcal{T}$ matrix in
Eq.\eqref{eq:BS-1d-reg}, we consider the $\mathcal{V}$ matrix in the
helicity basis and represent it in momentum space, with momenta
discretized using Gaussian quadrature. The $\mathcal{T}$ matrix is
then obtained via the Haftel–Tabakin matrix inversion
method\cite{Haftel:1970zz}: 
\begin{align}
\mathcal{T} = \left(1-\mathcal{V}\tilde{\mathcal{G}}\right)^{-1}
\mathcal{V}.
\end{align}
The $\mathcal{T}$ matrix in the helicity basis does not possess
definite parity. To determine the spin-parity classification of the
$P_{c\bar{c}sss}$ states, we transform the transition amplitudes into
partial-wave amplitudes with well-defined parity: 
\begin{align}
\mathcal{T}^{J\pm}_{\lambda'\lambda} =
\frac{1}{2}\left[\mathcal{T}^{J}_{\lambda'\lambda} \pm
\eta_1\eta_2(-1)^{s_1+s_2+\frac{1}{2}}
\mathcal{T}^{J}_{\lambda',-\lambda}\right],
\end{align}
where $\mathcal{T}^{J\pm}$ denotes the partial-wave transition
amplitude with total angular momentum $J$ and parity
$(-1)^{J\pm1/2}$. The factor $1/2$ ensures that no additional scaling
is needed when reverting to the partial-wave component: 
\begin{align}
\mathcal{T}^{J}_{\lambda'\lambda} =
\mathcal{T}^{J+}_{\lambda'\lambda} +\mathcal{T}^{J-}_{\lambda'\lambda}.
\end{align}

To analyze the dynamically generated $P_{c\bar{c}sss}$ resonances, we
reformulate the $\mathcal{T}$ matrix in the $IJL$ particle
basis~\cite{Machleidt:1987hj}. The relations between the
$\mathcal{T}$-matrix elements in these two bases are given by 
\begin{align}
\mathcal{T}^{JS'S}_{L'L} = \frac{\sqrt{(2L+1)(2L'+1)}}{2J+1}
\sum_{\lambda'_1\lambda'_2\lambda_1\lambda_2}
\left(L'0S'\lambda'|J\lambda'\right)
\left(s'_1\lambda'_1s'_2-\lambda'_2|S'\lambda'\right)
\left(L0S\lambda|J\lambda\right)
\left(s_1\lambda_1s_2-\lambda_2|S\lambda\right)
\mathcal{T}^{J}_{\lambda'_1\lambda'_2,\lambda_1\lambda_2}.
\end{align}
In the present analysis, we focus on the diagonal components of
$\mathcal{T}^{JS}_{L}$, as these provide the dominant contributions to
the production of triple-strangeness hidden-charm pentaquarks. 

\section{Results and discussions \label{sec:3}}
\begin{figure}[htbp]
  \centering
  \includegraphics[scale=0.51]{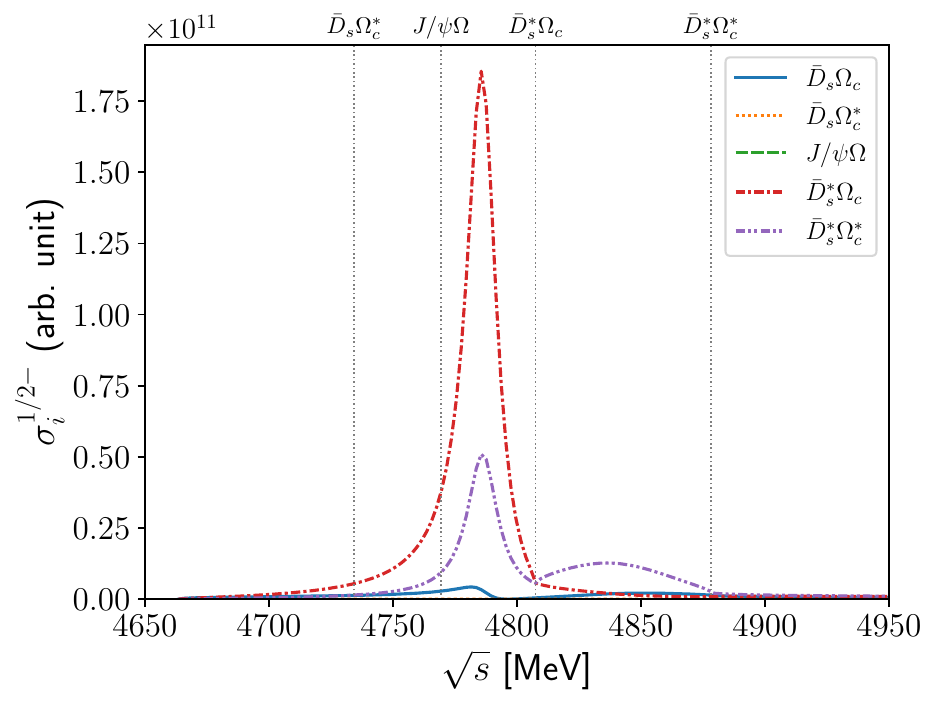}
  \includegraphics[scale=0.51]{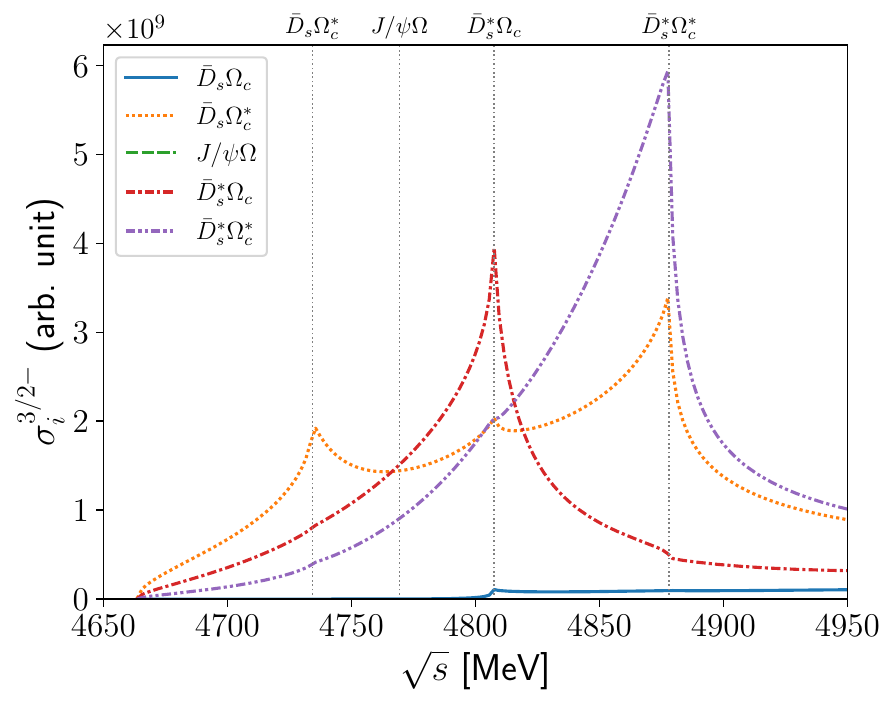}
  \includegraphics[scale=0.51]{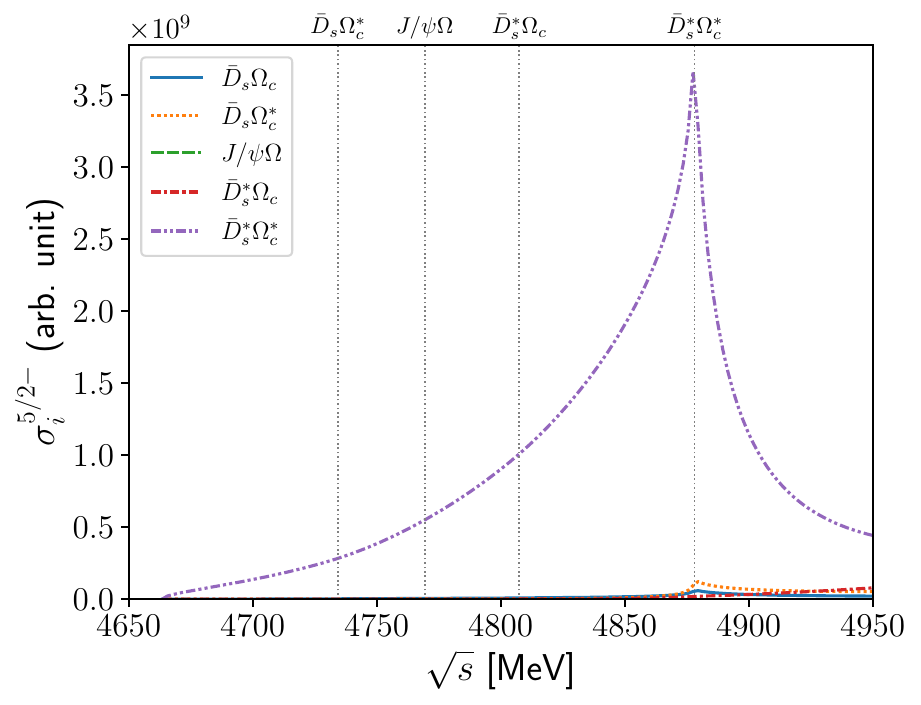}
  \includegraphics[scale=0.51]{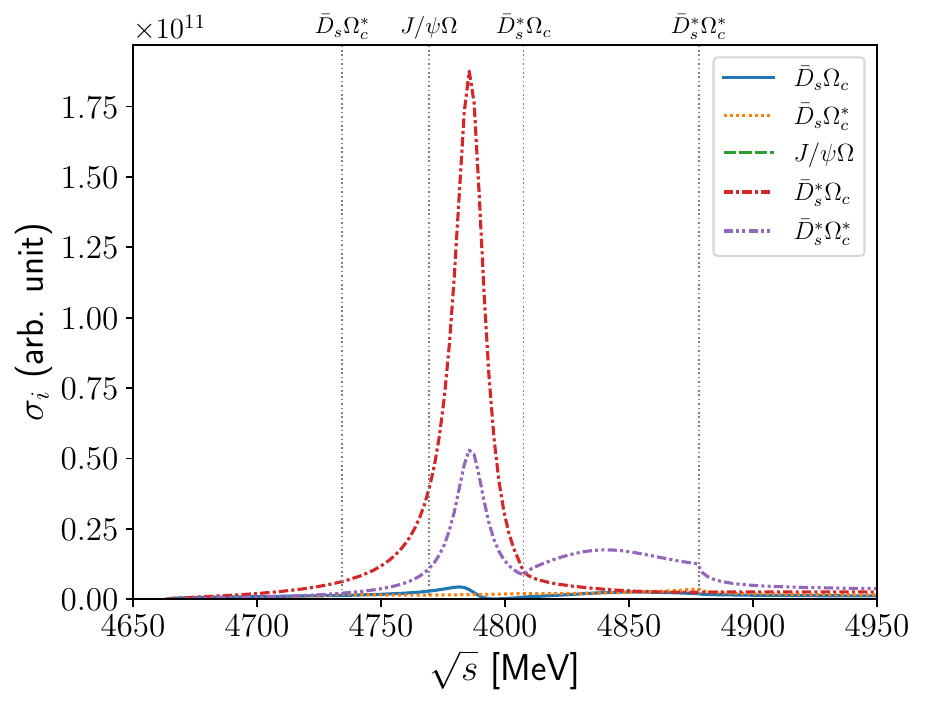}
  \caption{Elastic partial-wave cross sections for
    $J^P=1/2^-,\,3/2^-,\, 5/2^-$ (upper left, upper right, and lower
    left panels) and total cross section (lower right panel) versus
    total energy. The subscript $i$ labels the two-particle states
    shown in the legend.} 
  \label{fig:4} 
\end{figure}
In Fig.~\ref{fig:4}, we show the numerical results for the elastic
partial-wave cross sections for the given spin-parity classifications
$J^P=1/2^-,\,3/2^-,\, 5/2^-$, expressed as
\begin{align}
\sigma^{J\pm} = \frac{2J+1}{(2s_1+1)(2s_2+1)}
  \sum_{\lambda'\lambda}\mathrm{p}'  
  \left|\mathcal{T}_{\lambda'\lambda}^{J\pm}\right|^2,   
\end{align}
where $s_1$ and $s_2$ denote the spins of the incoming meson and
baryon, respectively.
In the upper left panel of Fig.~\ref{fig:4}, a peak clearly emerges in
the $\bar{D}_s^*\Omega_c$ channel, which can be identified as the
first triple-strangeness hidden-charm pentaquark resonance,
$P_{c\bar{c}sss}(4787)$ with $J^P=1/2^-$. This peak also appears in the
$\bar{D}_s^*\Omega_c^*$ channel. On the other hand, we observe a tiny
effect from the $\bar{D}_s\Omega_c$ channel, and we do not see any
signals in the $\bar{D}_s\Omega_c^*$ and $J/\psi \Omega$ channels. 
Note that the broad resonance structure is found in the
$\bar{D}_s^*\Omega_c^*$ channel, which is related to the 
second resonance $P_{c\bar{c}sss}(4841)$ with
$J^P=1/2^-$. Interestingly, the $P_{c\bar{c}sss}(4841)$ state does not
appear in other channels. We will discuss the reason later when presenting the results for the coupling strengths to the relevant channels. The right upper panel of Fig.~\ref{fig:4} draws the results
for the $\sigma_i^{3/2^-}$. While we do not observe any resonances, we
see three cusp structures at the thresholds corresponding to the
$\bar{D}_s\Omega_c^*$, $\bar{D}_s^*\Omega_c$, and
$\bar{D}_s^*\Omega_c^*$ channels. It is interesting to see that the
$\bar{D}_s\Omega_c^*$ channel exhibits the three cusp structures even
at the $\bar{D}_s^*\Omega_c$ and $\bar{D}_s^*\Omega_c^*$ thresholds. 
In the case of the $J^P=5/2^-$ partial wave, which is depicted in the
lower left panel, the single cusp structure is revealed only at the
$\bar{D}_s^*\Omega_c^*$ threshold. This is natural, since it is only
possible to have the total spin $5/2$ from the $\bar{D}_s^*\Omega_c^*$
channel in $S$-wave. Finally, we plot the total elastic cross section
including the relevant partial waves with both negative and positive
parities. We clearly observe the $P_{c\bar{c}sss}(4787)$ state in the
$\bar{D}_s^*\Omega_c$ channel, and both the $P_{c\bar{c}sss}(4787)$
and $P_{c\bar{c}sss}(4841)$ in the $\bar{D}_s^*\Omega_c^*$ channel.
In constrast to the cases of the $P_{c\bar{c}}$, $P_{c\bar{c}s}$, and
$P_{c\bar{c}ss}$, we do not find any $P_{c\bar{c}sss}$ resonances with
positive parity ($P$-wave). 

\begin{figure}[htbp]
  \centering
  \includegraphics[scale=0.55]{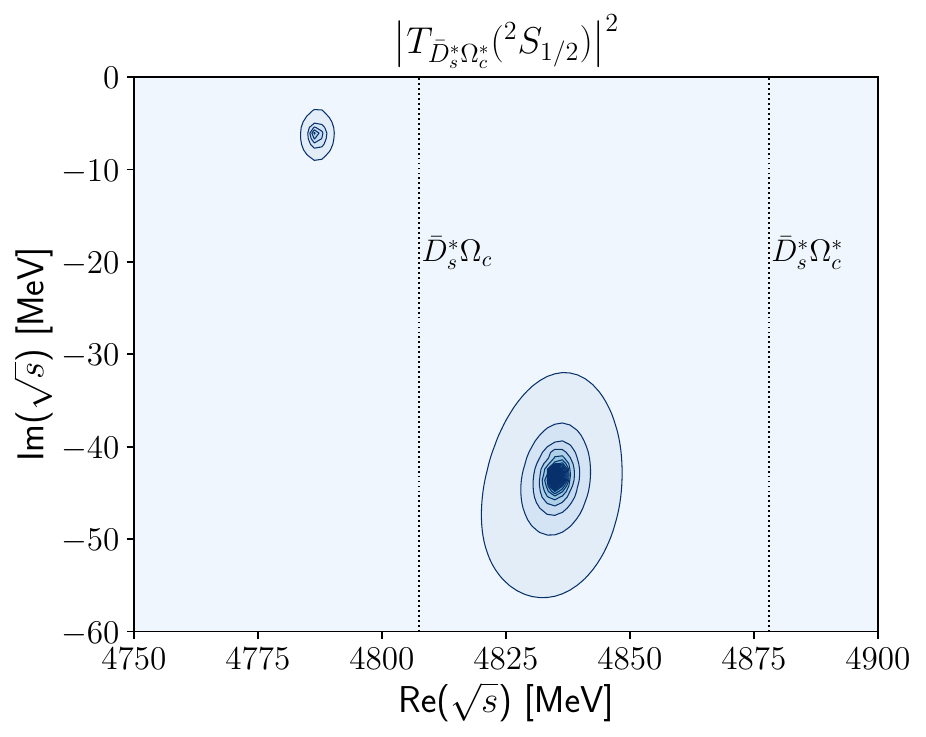}
  \caption{Contour plot of the modulus squared for the
    $\bar{D}_s^*\Omega_c^*$ transition amplitude in the complex total 
    energy plane. Two poles are observed, which correspond to the
    $P_{c\bar{c}sss}(4787)$ (upper left pole) and $P_{c\bar{c}sss}(4841)$
    (lower right pole). }  
  \label{fig:5} 
\end{figure}
Figure~\ref{fig:5} shows the contour plot of the modulus squared for
the  $\bar{D}_s^*\Omega_c^*$ transition amplitude,
$|T_{\bar{D}_s^*\Omega_c^*}|^2$ in the complex total energy plane. We
can easily find the pole positions for the two triple-strangeness
hidden-charm pentaquarks states. The first resonance,
$P_{c\bar{c}sss}(4787)$, is located at $(4786.9-i7.2)$ MeV below the
$\bar{D}_s^*\Omega_c$ threshold. It indicates that the width of
$P_{c\bar{c}sss}(4787)$ is $\Gamma=14.4$ MeV. The
$P_{c\bar{c}sss}(4841)$ state is positioned at $(4840.5-i43.1)$ MeV
below the $\bar{D}_s^*\Omega_c^*$ threshold. 
It has a larger width compared to the first resonance, i.e., $\Gamma=86.2$
MeV.  

\begin{table}
  \caption{\label{tab:2}Coupling strengths of the two $P_{c\bar{c}sss}$'s
      with $J^P=1/2^-$ to all channels involved.}  
   \centering
   \begin{tabular*}{\linewidth}{@{\extracolsep{\fill}} lcc}
      \toprule
    & $P_{c\bar{c}sss}(4787)$ & $P_{c\bar{c}sss}(4841)$ \\
    $\sqrt{s_R}$[MeV] & $4786.9-i7.2$ & $4840.5-i43.1$ \\\hline
   $g_{\bar{D}_s \Omega_c({}^2S_{1/2})}$            & $3.40+i2.26$    & $6.71-i3.58$  \\
   $g_{\bar{D}_s \Omega_c^*({}^4D_{1/2})}$          & $0.40+i0.11$    & $1.71-i1.22$  \\
   $g_{\bar{D}_s^* \Omega_c({}^2S_{1/2})}$          & $-13.23-i2.24$  & $1.61+i5.79$  \\
   $g_{\bar{D}_s^* \Omega_c({}^4D_{1/2})}$          & $-0.03-i0.02$   & $-1.13-i0.43$  \\
   $g_{J/\psi \Omega({}^2S_{1/2})}$        & $-0.03-i0.00$  & $0.02+i0.01$  \\
   $g_{J/\psi \Omega({}^4D_{1/2})}$        & $-0.00+i0.00$    & $0.00+i0.00$  \\
   $g_{J/\psi \Omega({}^6D_{1/2})}$        & $0.00-i0.00$    & $0.00+i0.01$  \\
   $g_{\bar{D}_s^* \Omega_c^*({}^2S_{1/2})}$        & $-10.19+i2.86$  & $20.41+i3.50$  \\
   $g_{\bar{D}_s^* \Omega_c^*({}^4D_{1/2})}$        & $0.00-i0.03$    & $-0.07-i0.07$  \\
   $g_{\bar{D}_s^* \Omega_c^*({}^6D_{1/2})}$        & $0.29+i0.01$    & $0.04+i0.18$  \\
         \bottomrule
   \end{tabular*}
 \end{table}
The coupling strength can be extracted from the residue of the
partial-wave transition amplitude, which is expressed as 
\begin{align}
  \mathcal{R}_{ab} =\lim_{s\to s_R} (s-s_R)\,\mathcal{T}_{ab}/4\pi = g_ag_b.
  \label{eq:residue}
\end{align}
We want emphasize that the definition of the coupling strength in
Eq.~\eqref{eq:residue} does not allow one to extract its absolute
sign. To determine the relative signs, thus, we consider the real
part of the coupling to the lowest threshold channel to be positive. 

Table~\ref{tab:2} lists the values of the coupling strengths for the
$P_{c\bar{c}sss}(4787)$ and $P_{c\bar{c}sss}(4841)$ to all channels
involved, in units of GeV. As shown in the upper left panel of
Fig.~\ref{fig:4}, the hidden-charm pentaquark $P_{c\bar{c}sss}(4787)$
state appears in the $\bar{D}_s^*\Omega_c$ and $\bar{D}_s^*
\Omega_c^*$ channels. This indicates that these two channels are the
most dominant ones. As expected, the coupling strength 
$g_{\bar{D}_s^*\Omega_c}({}^2S_{1/2})$ dominates over all other
coupling strengths, followed by
$g_{\bar{D}_s^*\Omega_c^*}({}^2S_{1/2})$. The mixing with the $D$-wave
is negligibly small. The coupling strength to the $\bar{D}_s
\Omega_c$ channel is marginal compared with those to the
$\bar{D}_s^*\Omega_c$ and $\bar{D}_s^*\Omega_c^*$ channels, as already 
illustrated in the upper left panel of Fig.~\ref{fig:4}.

The $P_{c\bar{c}sss}(4841)$ is only found in the
$\bar{D}_s^*\Omega_c^*$ channel. The corresponding coupling strength
is $g_{\bar{D}_s^*\Omega_c^*}({}^2S_{1/2})=(20.41+i3.50)$~GeV, the largest among all channels. Although it is not possible to conclude whether 
these two triple-strangeness hidden-charm resonances are 
genuine pentaquarks or molecular states solely from the
coupling strengths, they do provide partial insight into the nature of
these resonances. Examining the results in Table~\ref{tab:2}, the
$P_{c\bar{c}sss}(4787)$ is best interpreted as a mixed state of the
$\bar{D}_s^*\Omega_c$ and $\bar{D}_s^*\Omega_c^*$ channels, whereas
the second resonance, $P_{c\bar{c}sss}(4841)$, is most likely a
$\bar{D}_s^*\Omega_c^*$ molecular state.
  
Since several theoretical predictions for the pentaquark states with
$S=-3$ already exist, it is worthwhile to compare them with the
present results. Meng et al.~\cite{Meng:2019fan} studied the
$P_{c\bar{c}sss}$ states as compact pentaquarks within a quark
potential model. They found seven pentaquark states with spin-parity
$J^P=1/2^-$. The mass of the lowest $1/2^-$ state in
Ref.~\cite{Meng:2019fan} is 4855~MeV, which is close to the second
resonance $P_{c\bar{c}sss}(4841)$ observed in the present work. In
addition, they predicted nine $P_{c\bar{c}sss}$ states with $J^P=3/2^-$
and fifteen different pentaquarks with $J^P=5/2^-$. Even more pentaquark
states with positive parity were also predicted in
Ref.~\cite{Meng:2019fan}. 
Wang et al.~\cite{Wang:2021hql} constructed the hadronic potentials
based on the one-boson-exchange model and four different coupled
channels: $\bar{D}_s\Omega_c$, $\bar{D}_s\Omega_c^*$, 
$\bar{D}_s^*\Omega_c$, and $\bar{D}_s^*\Omega_c^*$. They then solved
the coupled-channel Schr\"odinger equation in the presence of these
hadronic potentials. They suggested two triple-strangeness
hidden-charm pentaquark states with $J^P=3/2^-$ and $5/2^-$ as bound
states of the $\bar{D}_s^*\Omega_c$ and $\bar{D}_s^*\Omega_c^*$
channels, rather than as resonances. To describe the decays of these
pentaquark states, they additionally employed the quark interchange
model. Roca et al.~\cite{Roca:2024nsi} also investigated the
$P_{c\bar{c}sss}$ states. However, they concluded that the
meson--baryon interaction they employed was insufficient to generate a
bound or resonance state. In the present work, on the other hand,
we have further considered the $\bar{D}_s\Omega_c$,
$\bar{D}_s^*\Omega_c$, and $\bar{D}_s^*\Omega_c^*$ channels, which are
absent in Ref.~\cite{Roca:2024nsi}. We have also included the  
scalar-isoscalar meson-exchange diagrams. It is essential to take
these contributions into account in order to generate the
$P_{c\bar{c}sss}$ states.

\begin{figure}[htbp]
  \centering
  \includegraphics[scale=0.55]{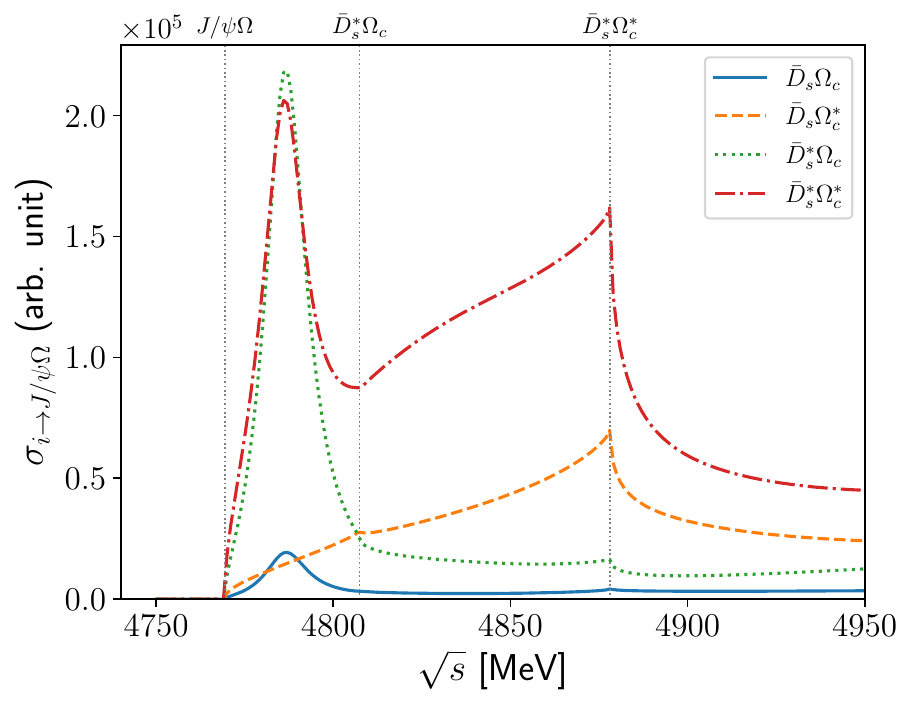}
  \caption{Total cross section as a function of energy for the transitions from the
$\bar{D}_s^{(*)} \Omega_c^{(*)}$ channels to the $J/\psi\Omega$
channel. The subscript $i$ labels the two-particle states
    shown in the legend.} 
  \label{fig:6} 
\end{figure}
The $P_{c\bar{c}}$ and $P_{c\bar{c}s}$ were experimentally found to
decay into the $J/\psi N$ and $J/\psi \Lambda$,
respectively~\cite{LHCb:2015yax, LHCb:2019kea, LHCb:2020jpq,
  LHCb:2021chn, LHCb:2022ogu}. So, we anticipate that the
hidden-charm pentaquark states with triple-strangeness may decay into
$J/\psi$ and $\Omega$, if they are observed experimentally. Therefore,
it is worthwhile to evaluate the transition cross sections from the
$\bar{D}_s^{(*)} \Omega_c^{(*)}$ channels to the $J/\psi \Omega$
channel. Figure~\ref{fig:6} shows the numerical results for the total
transition cross sections of $\bar{D}_s^{(*)} \Omega_c^{(*)}\to
J/\psi\Omega$. We can evidently observe the first resonance
$P_{c\bar{c}sss}(4787)$ in the $J/\psi \Omega$ invariant mass
spectrum. Interestingly, we can also see the resonance structure in
the $\bar{D}_s\Omega_c$ channel, though its strength is weaker than in
the $\bar{D}_s^*\Omega_c$ and $\bar{D}_s^*\Omega_c^*$ channels.
Since the $\bar{D}_s\Omega_c^*$ channel yields only the $D$-wave
contribution to the $\bar{D}_s\Omega_c^*\to J\psi \Omega$ interaction,
we do not find any signals for the resonance. 
On the other hand, the second resonance $P_{c\bar{c}sss}(4841)$ is
screened by the cusp structures and background contributions. So, we
expect that only the first resonance could be observed experimentally.

\section{Summary and conclusions\label{sec:4}}
We have investigated the possible existence of the triple-strangeness
hidden-charm pentaquark states in the off-shell coupled-channel
formalism. We have considered the open-charm meson-baryon channels 
$\bar{D}_s\Omega_c$, $\bar{D}_s\Omega_c^*$, $\bar{D}_s^*\Omega_c$, and
$\bar{D}_s^*\Omega_c^*$, as well as the hidden-charm channels 
$J/\psi\Omega$. We have constructed the two-body kernel Feynman
amplitudes using the effective Lagrangian based on the 
hidden local symmetry, constrained by heavy-quark spin
symmetry. We have solved the coupled Blankenbecler-Sugar equation in
the partial-wave helicity basis and have searched for poles in the
complex energy plane to identify possible resonance states.   

We have found two triple-strangeness hidden-charm pentaquark states 
with $J^P=1/2^-$: $P_{c\bar{c}sss}(4787)$ and
$P_{c\bar{c}sss}(4841)$. The $P_{c\bar{c}sss}(4787)$ couples
dominantly to the $\bar{D}_s^*\Omega_c$ and $\bar{D}_s^*\Omega_c^*$
channels, while the $P_{c\bar{c}sss}(4841)$ couples almost exclusively
to the $\bar{D}_s^*\Omega_c^*$ channel. We have also provided the
coupling strengths of these states to all relevant channels. Examining
the results for the coupling strengths, the $P_{c\bar{c}sss}(4787)$ is
best interpreted as a mixed state of the $\bar{D}_s^*\Omega_c$ and
$\bar{D}_s^*\Omega_c^*$ channels, whereas the second resonance,
$P_{c\bar{c}sss}(4841)$, is most likely a $\bar{D}_s^*\Omega_c^*$
molecular state. Finally, we have evaluated the total
transition cross sections of $\bar{D}_s^{(*)} \Omega_c^{(*)}\to
J/\psi\Omega$. The first resonance $P_{c\bar{c}sss}(4787)$ was clearly
observed in the $J/\psi \Omega$ invariant mass spectrum, whereas the
second one $P_{c\bar{c}sss}(4841)$ was screened by the cusp structures
and background contributions. Therefore, we expect that only the first
resonance can be observed experimentally.  

\begin{acknowledgments}
The present work was supported by the Young Scientist Training (YST)
Program at the Asia Pacific Center for Theoretical Physics (APCTP)
through the Science and Technology Promotion Fund and Lottery Fund of 
the Korean Government and also by the Korean Local Governments –
Gyeongsangbuk-do Province and Pohang City (SC), the Basic Science
Research Program through the National Research Foundation of Korea
funded by the Korean government (Ministry of Education, Science and
Technology, MEST), Grant-No. RS-2025-00513982 (HChK), and the PUTI Q1  
Grant from University of Indonesia under contract
No. PKS-206/UN2.RST/HKP.05.00/2025 (TM).
\end{acknowledgments}

\bibliography{Pcsss}
\bibliographystyle{apsrev4-2}

\end{document}